\documentclass[preprint,showkeys,aps,prb]{revtex4}                                                             
\usepackage[english]{babel}

\usepackage[dvips]{graphicx}
\usepackage{amssymb}

\begin{document}

\title{Ergodicity of a Singly-Thermostated Harmonic Oscillator}

\author{
William Graham Hoover, Ruby Valley Research Institute,                  \\
Highway Contract 60, Box 601 Ruby Valley, Nevada 89833, USA             \\
Julien Clinton Sprott, Department of Physics,                           \\
University of Wisconsin - Madison, Wisconsin 53706, USA                 \\
Carol Griswold Hoover, Ruby Valley Research Institute,                  \\
Highway Contract 60, Box 601 Ruby Valley, Nevada 89833, USA             \\
}

\date{today}



\vspace{0.1cm}

\begin{abstract}
Although Nos\'e's thermostated mechanics is formally consistent with Gibbs' canonical
ensemble, the thermostated Nos\'e-Hoover ( harmonic ) oscillator, with its mean kinetic
temperature controlled, is far from ergodic. Much of its phase space is occupied
by regular conservative tori.  Oscillator ergodicity has previously been achieved
by controlling two oscillator moments with {\it two} thermostat variables.  Here
we use computerized searches in conjunction with visualization to find
{\it singly}-thermostated motion equations for the oscillator which are consistent with
Gibbs' canonical distribution.  Such models are the simplest able to bridge the gap
between Gibbs' statistical ensembles and Newtonian single-particle dynamics.
\end{abstract}

\maketitle

\section{Ergodicity of the Equations of Motion}

Gibbs' statistical mechanics is based on summing contributions from ensembles
of similar systems.  His {\it microcanonical} ensemble includes all the states of
a given system which have the same energy.  These energy states are accessible to
a single ``ergodic'' system obeying Newtonian mechanics\cite{b1}.  A periodic
hard-disk or hard-sphere fluid is the usual example. Comparisons of Monte Carlo
microcanonical-ensemble averages with molecular dynamics dynamical averages have
confirmed this equivalence, even for small systems of just a few particles\cite{b2}.

Certainly Boltzmann and Gibbs both realized that {\it all} states need to be accessible to
the dynamics in order for the dynamical and phase averages to correspond.  The Ehrenfests
had a practical definition of ``quasiergodicity''.  They used the word to indicate that the
dynamics eventually comes ``arbitrarily close'' to all states. Their idea expresses very well
our own view of what we call ``ergodicity'' in the present work.

Gibbs' {\it canonical} ensemble sums Boltzmann-weighted contributions from
{\it all} energy states.  The underlying idea is that the system of interest is
weakly coupled to a heat reservoir with an ideal-gas density of states
characteristic of a fixed kinetic temperature $T$ .  Shuichi Nos\'e\cite{b3,b4}
developed a dynamics consistent with the canonical distribution by including a
``time-scaling'' variable $s$ and its conjugate momentum $p_s$ in the equations of
motion.  The new momentum $p_s$ acts as a thermostat variable capable of exchanging
energy between the system and a heat reservoir at temperature $T$.

Hoover showed that a harmonic oscillator thermostated in this way is not at all
ergodic\cite{b5}.  That is, there is no initial condition from which the oscillator
is able to access {\it all} of its phase-space states.  Instead, this thermostated
oscillator has a nonergodic highly-complicated multi-part phase-space
structure\cite{b6}.  There are infinitely-many regular nonchaotic orbits embedded
in a single chaotic sea.  Where ``Chaos'' controls the motion two closeby points,
$r_1(t)$ and $r_2(t)$ ,
tend to separate from one another exponentially fast, either forward or backward
in time. Such a motion is said to be ``Lyapunov unstable''.  The averaged separation
rate is described by the largest Lyapunov exponent, $\lambda_1$ :
$$
\delta \equiv | \ r_2 - r_1 \ | \simeq e^{\lambda_1 t} \ ; \
\lambda_1 \equiv \langle \ \lambda_1(t) \ \rangle \ .
$$

The time-averaged Lyapunov exponent $\lambda_1$ is computed as an average of the
instantaneous local Lyapunov exponent, $\lambda_1(t)$.  The local value is only
rarely zero, even on conservative tori, where the long-time averages vanish.
We illustrate $\lambda_1(t)$ for the Nos\'e-Hoover
oscillator in {\bf Figure 1}.  We choose the simplest equations of motion,
$$
\{ \ \dot q = p \ ; \ \dot p = -q - \zeta p \ ; \ \dot \zeta = p^2 - 1 \ \}
\ [ \ {\rm NH} \ ] \ .
$$
They are time-reversible: any time-ordered sequence $\{ \ +q,+p,+\zeta \ \}$
satisfying the motion equations has a time-reversed backward twin,
$\{ \ +q,-p,-\zeta \ \}$ satisfying the same equations.  The Nos\'e-Hoover
oscillator in $\{ \ q,p,\zeta \ \}$ space is an improved and simplified version
of Nos\'e's dynamics, which occupies a four-dimensional $\{ \ q,p,s,p_s\ \}$
space\cite{b5,b6}.

For this Nos\'e-Hoover oscillator we have computed the local Lyapunov exponent
on a grid of about a million points by the simple
expedient of integrating backward in time and then forward, for a time of 100 in
both directions.  The ``reversed'' trajectory going forward in time can
be compared to a nearby constrained ``satellite'' trajectory. We compute
the instantaneous value of the time-dependent Lyapunov exponent just as
the $\zeta = 0$ plane is crossed.  In the Figure red corresponds to the
most positive exponent value and blue to the most negative.  Within the chaotic
sea the largest ( time-averaged ) Lyapunov exponent is $0.013_9$ .  See Reference
6 for details.

\begin{figure}
\includegraphics[width=4in,angle=-90.]{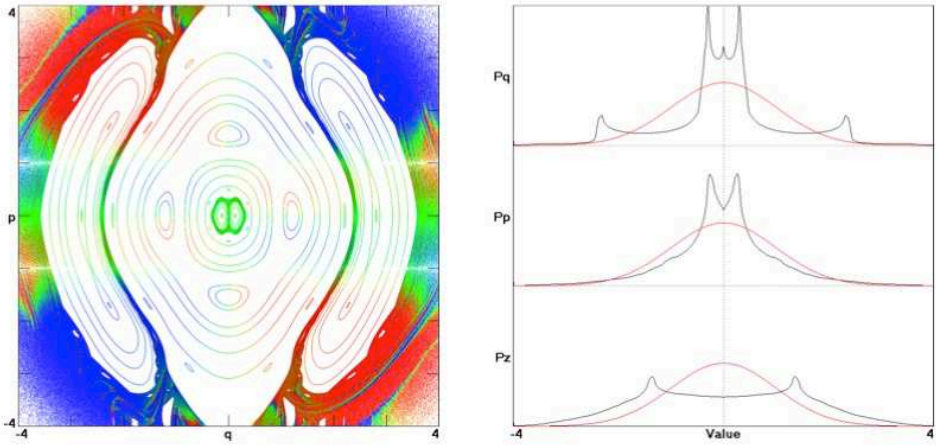}
\caption{
This Nos\'e-Hoover oscillator phase-space section corresponds to the plane
$\zeta = 0$ .  The coloring reflects the local value of the instantaneous
Lyapunov exponent at each $( \ q,p,0 \ )$ point, with red least stable and
blue most.  The distributions of $\{ \ q,p,\zeta \ \}$  in the chaotic sea
are compared to Gibbs' Gaussian distributions at the right. The white space
indicates nonchaotic regions filled with regular nested tori, some of which
are shown.  In the chaotic sea $\lambda_1 = \langle \ \lambda_1(t) \ \rangle
= 0.0139$ .
}
\end{figure}

Outside the chaotic sea lie an infinite number of regular orbits. All have
a largest time-averaged Lyapunov exponent ( and also a smallest ) of zero.
Because the oscillator is prototypical of systems with smooth minima in
their energy surfaces a considerable effort has been made to find motion
equations providing Gibbs' canonical distribution for
it\cite{b7,b8,b9,b10,b11,b12,b13,b14,b15,b16,b17,b18}.

\section{Feedback Control of Oscillator Moments}
For simplicity we choose units of force, mass, time, and temperature
corresponding to choosing the oscillator force constant, mass, angular
velocity, and Boltzmann's constant all equal to unity.  In these units
and without any thermostating the oscillator motion equation is
$\ddot q = \dot p = - q$ .  Because distribution functions for the
oscillator's displacement and momentum can be described in terms of their
moments $\langle \ q^{2m}p^{2n} \ \rangle$ , it was natural to control
oscillator force and velocity moments with feedback variables such as
$\zeta$ and $\xi$ :
$$
\{ \ \dot q = p - \xi q\ ; \ \dot p = -q - \zeta p \ \} \ .
$$
The time dependence of the friction coefficients $\zeta(t)$ and $\xi(t)$
can be arranged so as to control one or more of the oscillator moments :
$$
\dot \zeta = (p^2/T) - 1 \rightarrow \langle \ p^2 \ \rangle \equiv T \ ; \
\dot \xi = (q^4/T^2) - 3(q^2/T) \rightarrow \langle \ q^4 \ \rangle \equiv 3T\langle \ q^2 \ \rangle \ \dots \ .
$$
Bulgac and Kusnezov, along with their coworkers Bauer and Ju\cite{b15,b16}, considered a
variety of simple systems and concluded that {\it cubic} contributions to the control
equations, such as those in the [ HH ] and [ JB ] equations below,
were especially useful in promoting chaos and ergodicity.

Over thirty years dozens of investigators explored the ergodicity of thermostated
oscillators\cite{b7,b8,b9,b10,b11,b12,b13,b14,b15,b16,b17,b18}.  Three successful models,
the Hoover-Holian\cite{b13}, Ju-Bulgac\cite{b15}, and Martyna-Klein-Tuckerman models\cite{b17}
resulted.  With all of the thermostat relaxation times set equal to unity, these models have
the following forms :
$$
\dot q = p \ ; \ \dot p = -q -\zeta p - (\xi p^3/T) \ ; \
\dot \zeta = (p^2/T) - 1 \ ; \ \dot \xi = (p^4/T^2) - 3(p^2/T) \ ; \ {\rm [ \ HH \ ]}
$$
$$                                                                                                                                                             
\dot q = p \ ; \ \dot p = -q - \zeta^3 p - (\xi p^3/T) \ ; \                                                                                                          
\dot \zeta = (p^2/T) - 1 \ ; \ \dot \xi = (p^4/T^2) - 3(p^2/T) \ ; \ {\rm [ \ JB \ ]}                                                                                              
$$
$$                                                                                                                                                             
\dot q = p \ ; \ \dot p = -q -\zeta p  \ ; \                                                                                                          
\dot \zeta = (p^2/T) - 1 - \xi \zeta\ ; \ \dot \xi = \zeta^2 - 1 \ . \ {\rm [ \ MKT \ ]} 
$$
In all three cases the phase-space continuity equation,
$$
(\partial f/\partial t) = -(\partial f\dot q/\partial q) - (\partial f\dot p/\partial p) -
(\partial f\dot \zeta/\partial \zeta) - (\partial f\dot \xi/\partial \xi) \ .
$$   
shows that the motion equations are consistent with Gibbs' canonical distribution for the
$(q,p)$ coordinate-momentum pair :
$$
f(q,p) = e^{-q^2/2T}e^{-p^2/2T}/(2\pi T) \ .
$$
The distributions of the control variables $\zeta$ and $\xi$ are likewise Gaussians :
$$
f_{HH} = f_{MKT} \propto e^{-\zeta^2/2}e^{-\xi^2/2} \ ; \
f_{JB} \propto e^{-\zeta^4/4}e^{-\xi^2/2} \ .
$$
If the dynamics is ergodic, filling out the full distributions, these models 
reproduce Gibbs' canonical distribution.

Although there are no {\it theoretical} proofs that these dynamics obey the phase-space
distributions and relationships there is by now abundant {\it numerical} evidence
that the three two-thermostat approaches given above {\it are} ergodic\cite{b19}.
Recent work, particularly that of Patra and Bhattacharya\cite{b20,b21}, led us to search
for even simpler models, with three equations rather than four, for thermostated
oscillators.  We describe our own quite novel and successful findings next.

\section{Singly-Thermostated Ergodic Oscillator Models}

\begin{figure}
\includegraphics[width=4in,angle=-90.]{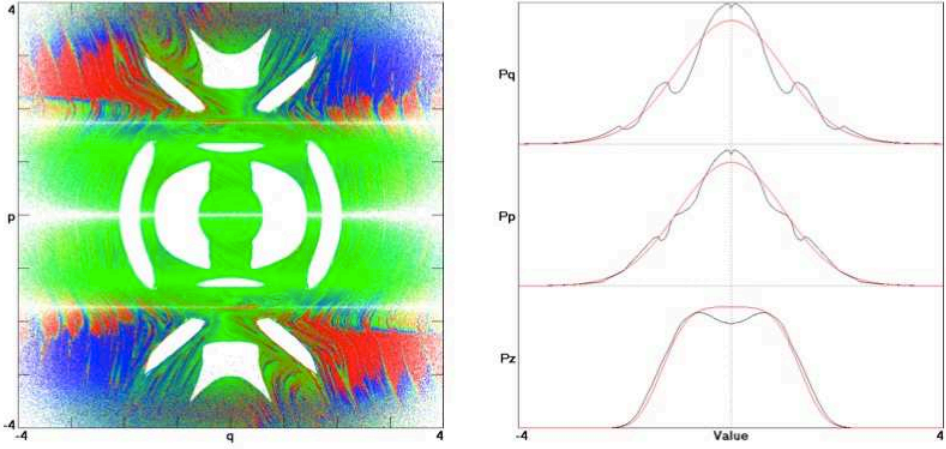}
\caption{
Single-thermostat cubic control enforcing the fourth-moment condition,
$\langle \ p^4 \ \rangle = 3\langle \ p^2 \ \rangle$ with $\alpha$
and $T$ both equal to unity and $\beta= 0$ . These choices show large
gaps in the cross section where the time-averaged Lyapunov exponent vanishes.
The probability densities within the chaotic sea are shown at the
right. The initial condition used here and in all succeeding figures
to access the chaotic sea is $( \ q,p,\zeta \ ) = ( \ 0,5,0 \ )$ .
The three horizontal nullclines at $\{ \ p = -\sqrt{3},0,+\sqrt{3} \ \}$
reflect the vanishing of the phase-space velocity component normal to
the $\zeta = 0$ plane.  $\lambda_1 = 0.1108$ in the chaotic sea.
}
\end{figure}

\begin{figure}
\includegraphics[width=4in,angle=-90.]{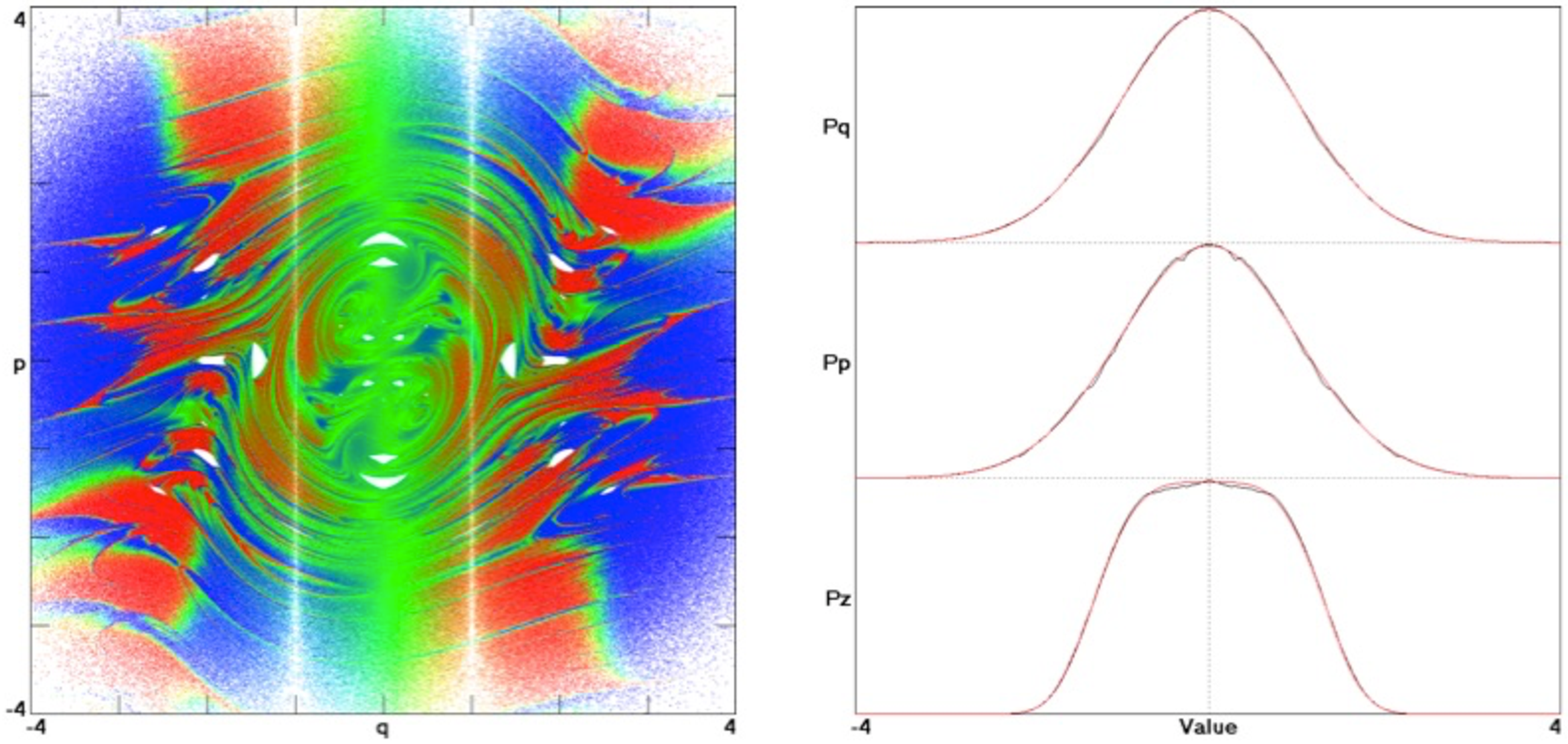}
\caption{
Single-thermostat cubic control enforcing the second moment condition
$\langle \ q^2 \ \rangle = 1$ with $\beta$ and $T$ set equal to unity
and $\alpha = 0$.  These choices show the presence of at least twenty
gaps ( or ``holes'' ) in the cross section where the friction coefficient
$\zeta$ vanishes.  The one-dimensional probability densities within the
chaotic sea are shown at the right along with the Gibbs' distributions
from the canonical ensemble. $\lambda_1 = 0.0905$ in the chaotic sea.
}
\end{figure}

A first look at the possibility of thermostating an oscillator with a single friction
coefficient corresponds to a variety of separate models.  We consider two of them
here. Both include cubic functions as suggested by Bulgac, Kusnezov, Ju, and Bauer.
The first oscillator model controls the fluctuation of the kinetic energy :
$$
\dot q = +p \ ; \ \dot p = -q - \alpha (\zeta^3p^3/T) \ ; \                                
\dot \zeta = \alpha [ \ (p^2/T)^2 - 3(p^2/T) \ ] \ .
$$
The second controls the fluctuation of the force :
$$
\dot q = +p - \beta \zeta^3q \ ; \ \dot p = -q \ ; \
\dot \zeta = \beta [ \ (q^2/T) - 1 \ ] \ .
$$
Neither of these approaches is successful.  In {\bf Figures 2 and 3} we show the
chaotic seas corresponding to these two models with first $\alpha$ and then $\beta$
equal to unity. In both cases we choose unit temperature, $T=1$ .  We plot $(q,p)$
points whenever the friction coefficient $\zeta$
changes sign.  These models both contain holes in the sea filled with regular
toroidal regions.  The one-dimensional distribution functions shown at the right
of these figures give an alternative view of the models' inability to describe
Gibbs' canonical distribution.  Both of these single-thermostat models are
failures.  In addition to plotting cross sections and probability densities one can
evaluate the likelihood of deviations from Gibbs' values as measured by the $\chi^2$
statistic described in Wikipedia, {\it Numerical Recipes}, and many other texts.

At least in retrospect it is natural to consider the possibility that a
{\it single} friction coefficient $\zeta$ might somehow control {\it two} moments
simultaneously, rather than just one. For example, consider simultaneous control of
{\it fluctuations} in {\it both} the force $\simeq \langle \ q^2 \ \rangle$
and the kinetic energy $\simeq \langle \ p^4 - 3p^2T \ \rangle$ :
$$
\dot q = +p - \beta \zeta^3q \ ; \ \dot p = -q -\alpha (\zeta^3p^3/T) \ ;
$$
$$
\dot \zeta = \beta [ \ (q^2/T) - 1 \ ] + \alpha [ \ (p^2/T)^2 - 3(p^2/T) \ ]
\ [ \ {\rm HS} \ ] \ .
$$
Our numerical work indicates that this [ HS ] ( Hoover-Sprott ) idea has merit.  With {\it two} free 
parameters $( \ \alpha,\beta \ )$ there are an infinite number of models which
could be tested against the predictions of Gibbs' canonical ensemble. This variety
could be extended further by including one or more relaxation times.  To choose
among the combinations of $\{ \  \alpha,\beta \ \}$ it is convenient to use computerized 
searches, either seeking minimum deviations with Monte Carlo searches\cite{b22} or by
choosing minima from grid-based arrays of $( \ \alpha,\beta \ )$ results.

\begin{figure}
\includegraphics[width=4in]{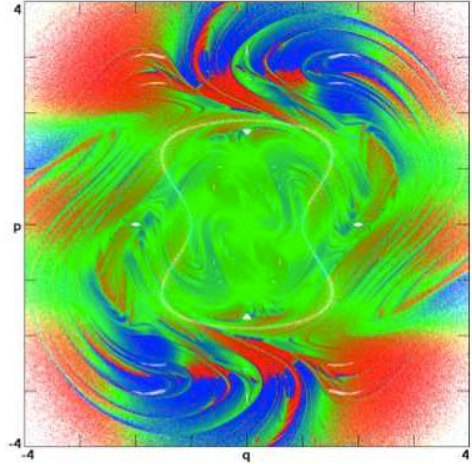}
\caption{
$( \ \alpha,\beta \ ) = ( \ 0.411,0.689 \ )$ .  A close inspection shows 26 prominent
``holes'' with a total measure less than one-half percent of the total. The Figure also
illustrates the typical figure-eight-shaped nullcline where the trajectory motion is
parallel to the $\zeta = 0$ plane.  Because the deviations of the various one-dimensional
probability densities are visually indistinguishable from Gibbs' canonical distributions
none of them is shown in Figures 4-6. Here $\lambda_1 = 0.1621$ in the chaotic sea.
}
\end{figure}

\begin{figure}
\includegraphics[width=4in]{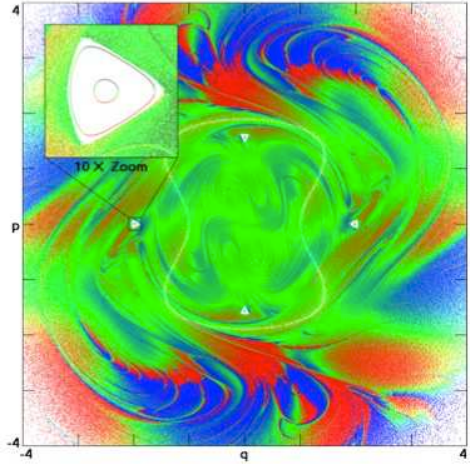}
\caption{
$( \ \alpha,\beta \ ) = ( \ 0.354,0.746 \ )$ likewise provides
``reasonable'' distributions and moments, but has four holes where nested tori
penetrate the cross section.  The tenfold ``zoom'' of one hole shows their
roughly triangular shape. Here $\lambda_1 = 0.1525$ in the chaotic sea.
}
\end{figure}

\begin{figure}
\includegraphics[width=4in,angle=-90.]{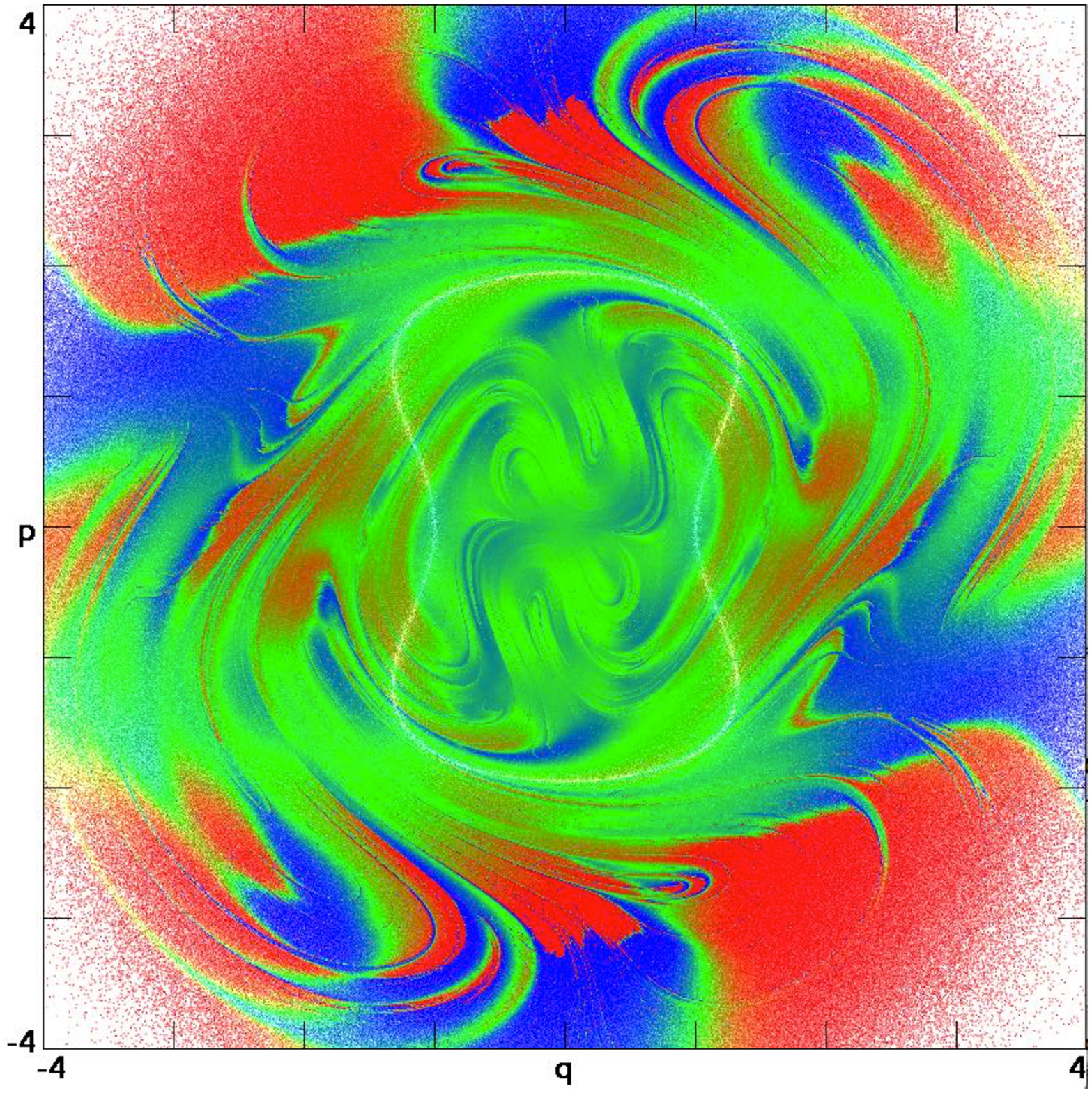}
\caption{
$( \ \alpha,\beta \ ) = ( \ 0.273,0.827 \ )$ is just one of an infinite number of
  combinations that is apparently ergodic. $\lambda_1 = 0.1450$ .
}
\end{figure}

We use standard Runge-Kutta integration methods throughout this work, fourth-order and fifth-order
as well as two types of ``adaptive'' integrators.  In the adaptive cases the timestep $dt$
is doubled if the error is ``small'' ( typically $10^{-16}$ ) and halved if the error is
``large'' ( typically $10^{-12}$ ).  The error estimate compares either fourth-order and
fifth-order results with the same timestep $dt$ or two fourth-order results with $dt$
and $dt/2$ .   There are no significant differences in the conclusions reached with any
of these several methods.  We carried out independent calculations in Nevada and in
Wisconsin.

With either two-parameter approach, the computation of moments is straightforward.
Optimizing the dynamics so as to seek out the Gibbs distribution can be accomplished
by evaluating either [ 1 ] moments of the distribution $f(q,p)$ or [ 2 ] values of
the distribution itself.  To follow the first approach we  minimized the summed-up
squared deviations of the first five nonvanishing Gibbsian moments :
$$
\sigma^2 \equiv [ \ (q^4/T^2) - 3 \ ] +[ \ (q^2p^2/T^2) - 1 \ ] +[ \ (p^4/T^2) - 3 \ ] +
$$
$$
[ \ (q^2/T) - 1 \ ] +[ \ (p^2/T) - 1 \ ] \ . 
$$

We evaluated $\sigma^2$ for thousands of runs.  Each used 100 million timesteps for
a particular pair of candidate values $( \ \alpha,\beta \ )$. Time-averaged values
of $\sigma^2$ suggest the range $0.25 < \alpha < 0.65$ with $1 < \alpha + \beta < 1.1$ . 
{\bf Figures 4-6} shows three typical cross-sectional plots of $\{ \ q,p \ \}$ sections
selected in this way. {\bf Figure 4}, with $( \ \alpha,\beta \ ) = ( \ 0.411,0.689 \ )$ ,
shows 26 noticeable ``holes'' with a total measure near one percent of the total.
The Figure also illustrates the figure-eight-shaped nullcline where trajectories
move parallel to the $\zeta = 0$ plane with $\dot \zeta \equiv 0$ ;
$$
\beta [ \ (q^2/T) - 1 \ ] + \alpha [ \ (p^2/T)^2 - 3(p^2/T) \ ] \equiv 0 \ .
$$
Despite all the holes indicating nested tori the distribution functions and their
first several moments are close to the Gibbsian ergodic values.  Evidently there is
no real substitute for looking at the sections themselves.

{\bf Figure 5}, with $( \ \alpha,\beta \ ) = ( \ 0.354,0.746 \ )$ , likewise provides
``reasonable'' distributions and moments, but has four holes where nested tori
penetrate the cross section.  Figures 4 and 5 hint at the extensive zoo of
topologies hidden in the $( \ \alpha,\beta \ )$ plane.  There are in addition patches
of values which evidently correspond to ergodicity.  Two examples which we think are
likely ergodic are :
$$
( \ 0.273,0.827 \ ) \ {\rm and} \ ( \ 0.274,0.826 \ ) \ .
$$
The first of these is plotted in {\bf Figure 6} .  It has no noticeable ``holes'',
indicating ergodicity within an accuracy of about one part per hundred thousand.
Again the Figure-Eight-Shaped white space indicates the nullcline, which depends
weakly on the precise value of the ratio $(\alpha/\beta)$ .

One can only search for ``holes'' in sections visually.  An example, which we thought
to be ergodic after a cursory inspection, is the combination
$( \ \alpha,\beta \ ) = ( \ 0.495,0.555 \ )$ , not shown here because visualizing
the holes requires magnification.  A close inspection of the $\zeta = 0$ section
reveals 36 tiny holes (!) corresponding to a single thin set of nested tori, including two near $( \ q,p \ ) = ( \ \pm 1.5,0.0 \ )$ .  These tori cross the $\zeta = 0$ plane in
36 separate places.

Because the numerical value of the summed squared-moment errors depends upon both
the initial conditions and the length of the trajectories a reasonable procedure
is to investigate visually, as second and third criteria for ergodicity, the
distributions themselves as well as their $(q,p)$ cross sections.  Such inspections reveal
the $( \  \alpha,\beta \ )$ pairs most promising from the standpoint of ergodicity.
The distributions found for the cross sections of the Figures 1-3 are included in
those Figures. From the visual standpoint such histograms show no significant
deviations from the ergodic distribution in the data displayed for figures 4 and 5,
despite the clear nonergodicity seen in the cross sections.  Our results suggest
overall that a {\it visual} inspection of two-dimensional cross sections is the most
reliable way to identify ergodicity in these three-dimensional dynamical systems.

An alternative method for evaluating ergodicity is to compute the probability that a
measured distribution of data points, such as $\{ \ q_i \ \}$ or $\{ \ p_i \ \}$
or $\{ \ \zeta_i \ \}$ comes from the expected Gaussian distribution of such points.
Comparing the probabilities for the three choices demonstrates the accuracy of
such a test.  So long as the sampling bins contain many points the mean-squared
deviation of the bin populations should be approximately equal to Gibbs' value. Such
tests implementing $\chi^2$ criteria can serve as useful indicators for deviations
from ergodicity.  In any doubtful case visual inspection is the only reliable
criterion.

\section{Phase-Space Density Flows}
An apparent alternative to solving the motion equations
$\{ \ \dot q,\dot p,\dot \zeta \ \} = v$ for a specimen oscillator is the solution of
Liouville's continuity equation, $\dot f/f = -\nabla \cdot v$ , so as to study the
details of the convergence ( or lack of it ) to Gibbs' canonical distribution.  We
briefly considered this approach and developed a straightforward finite-difference
program simulating the three-dimensional flow of the probability density
$f(q,p,\zeta,t)$ .  This program quickly led to negative densities.  A conservative
approach, passing probabilities between adjacent cells, can be implemented with a
swarm of $N$ moving particles, all of equal probability.  The instantaneous summed-up
density of these particles at any point in phase space can be made continuous and
twice-differentiable by defining and computing a smooth-particle density. This idea
is simplest to implement using Lucy's weight function\cite{b23} $w(r<h)$ with a
range $h$ of order two or three times the nearest-neighbor particle spacing:
$$
f(q,p,\zeta) \equiv \sum_i^N w(r-r_i) \ ; \ w(z=r/h<1) \propto (1 + 3 z)(1-z)^3 \ .
$$
We explored this idea using an initial condition $f(0 < q,p,\zeta < 1) \equiv 1$ 
and noticed that such a localized  initial value requires several
Lyapunov times to smooth out.  The particulate basis of the density guarantees that
there is no tendency for this solution of the Liouville flow to stabilize.  The
time-reversible nature of the flow guarantees that a smooth stationary solution can
only be obtained by adding in a time-averaging step.  A detailed investigation of
these ideas is likely worthwhile in that the compact three-dimensional nature of
these flows makes visualization easy.

\section{Summary}
It appears highly likely that a single thermostat variable {\it is} enough to provide Gibbs'
canonical distribution for a thermostated harmonic oscillator.  This question has
stimulated a relatively complex and varied literature over a 30-year period.  The
mathematicians are content to prove nonergodicity\cite{b24}.  The computational
physicists, ourselves included, have been prone to give up on the possibility of
ergodicity with a single thermostat variable\cite{b25}.  Thus our finding that one
thermostat {\it is} enough was a pleasant surprise.  The $( \ \alpha,\beta \ )$
model detailed here seems likely to be the simplest smoothly deterministic, ergodic,
time-reversible, and chaotic system for which the phase-space distribution is exactly known.

\section{Acknowledgment} Bill and Carol Hoover particularly appreciate a recent extended
conversation with John Ramshaw concerning the feasibility of ergodicity studies.  We
have also all benefited from experience gleaned from Puneet Patra's and Baidurya
Bhattacharya's investigations of the ergodicity of the Martyna-Klein-Tuckerman and their
own multi-moment [ PB ] oscillator models.

\pagebreak

\end{document}